\documentclass[aps,preprint,amsfonts,amssymb,amsmath,%
tightenlines,nofootinbib,showpacs,showkeys]{revtex4}

\newcommand{\cC}{\mathcal{C}}
\newcommand{\cP}{\mathcal{P}}
\newcommand{\cT}{\mathcal{T}}
\newcommand{\fE}{\mathfrak{E}}
\newcommand{\fH}{\mathfrak{H}}
\newcommand{\fL}{\mathfrak{L}}
\newcommand{\fR}{\mathfrak{R}}
\newcommand{\fS}{\mathfrak{S}}
\newcommand{\bbC}{\mathbb{C}}
\newcommand{\Ker}{\operatorname{Ker}}

\begin{document}



%
%

\title{On Existence of a Biorthonormal Basis Composed of Eigenvectors of
 Non-Hermitian Operators}
\author{Toshiaki Tanaka}
\email{ttanaka@mail.tku.edu.tw}
\affiliation{Department of Physics, Tamkang University,\\
 Tamsui 25137, Taiwan, R.O.C.}


\begin{abstract}

We present a set of necessary conditions for the existence of
a biorthonormal basis composed of eigenvectors of non-Hermitian
operators. As an illustration, we examine these conditions in
the case of normal operators. We also provide a generalization
of the conditions which is applicable to non-diagonalizable
operators by considering not only eigenvectors but also all
root vectors.

\end{abstract}


\pacs{02.30.Sa; 02.30.Tb; 03.65.Db}
\keywords{Non-Hermitian Hamiltonians; Biorthonormal bases; 
 Completeness; Diagonalizability; Normal operators}




\maketitle

Self-adjointness (or Hermiticity\footnote{What physicists usually
call \emph{Hermitian} operators correspond to what mathematicians
usually call \emph{symmetric} operators, which is a less restrictive
concept than \emph{self-adjoint} operators.}) of operators
corresponding to physical observables
is well-known as one of the postulates of quantum theory.
Nevertheless, non-Hermitian Hamiltonians have been often employed
in various areas of physical applications. This is mostly due to
the fact that the self-adjointness is too restrictive for practical
needs to describe physical systems which have in particular
dissipative and unstable nature.
Several years ago, a new explosion of research activities on
non-Hermitian Hamiltonians was triggered by Ref.~\cite{BB98a}
where the authors introduced the new notion of $\cP\cT$ symmetry.
Later, Mostafazadeh attempted to reformulate $\cP\cT$-symmetric
theory within the concept of pseudo-Hermiticity~\cite{Mo02a}
and derived various properties of pseudo-Hermitian theory
mostly under the assumption of the existence of a biorthonormal
basis composed of eigenvectors (or, root vectors) of operators,
see Ref.~\cite{Mo05b} and references cited therein. In response
to these works, there has appeared an increasing number of papers
in which the same assumption has been employed.

To the best of our knowledge, the validity of the assumption
was first questioned in Ref.~\cite{KS04} where the authors recalled
an important caution for metric operators in quasi- and
pseudo-Hermitian theories, which originated in Ref.~\cite{SGH92}
and also applies to $\cC$ operators in $\cP\cT$-symmetric
theory~\cite{BBJ02,BBJ04a}, though the paper has not been duly
appreciated in the literature appeared after it until now.
Then, it was independently pointed out in Ref.~\cite{GT06} that
ascertaining whether the assumption is indeed satisfied by
a given operator would be far from trivial.

A naive derivation of a biorthonormal system composed of
eigenvectors of non-Hermitian operators already appeared in
the classic book~\cite{MF53}, and some attempts with much
mathematical care dates back to (at the latest) the late
1960s~\cite{FGWR66,Wo67}. In Ref.~\cite{Wo67}, for instance,
the authors considered such a Hamiltonian $H$ which satisfies that
(i) $H$ is dissipative, (ii) $H$ can be expressed as a sum of
a self-adjoint (unperturbed) part $H_{0}$ and a bounded
(interaction) part $H_{1}$, such that (iii)
$H_{1}^{1/2}(H_{0}-\lambda I)^{-1}H_{1}^{1/2}$ is compact
for every regular point $\lambda\in\rho(H_{0})$ of
$H_{0}$.\footnote{See any elementary textbook on functional
analysis for the precise definition of mathematical
concepts such as dissipative (bounded, compact) operators,
regular point, etc and the commonly used notation such as
kernel $\Ker A$, resolvent set $\rho(A)$, point spectrum
$\sigma_{p}(A)$ of an operator $A$, etc appeared in this paper
without explicit explanation.}
Later, however, another naive derivation of a biorthonormal
system under a quite loose condition reappeared~\cite{FM81}.
As far as the list of the references indicates, it was
this work that the author of Ref.~\cite{Mo02a} relied on for
the existence of a `complete biorthonormal basis'.

The aim of this paper is to present what kinds of
conditions are necessary for the existence of a complete
biorthonormal basis by clarifying subtleties which lurk
behind the naive derivation in Ref.~\cite{FM81}. Although
they would be rather well-known facts among mathematicians,
they would not be duly recognized by physicists as the situation
described above clearly indicates. For the later discussions,
we first introduce the following notation:
\begin{align}
\fS_{\lambda}(A)&=\bigcup_{n=0}^{\infty}\Ker\bigl(
 (A-\lambda I)^{n}\bigr),\\
\fE_{0}(A)&=\overline{\bigl\langle\Ker(A-\lambda I)\bigm|
 \lambda\in\sigma_{p}(A)\bigr\rangle},\\
\fE(A)&=\overline{\bigl\langle\fS_{\lambda}(A)\bigm| \lambda\in
 \sigma_{p}(A)\bigr\rangle}.
\end{align}
That is, $\fS_{\lambda}(A)$ is the root subspace spanned by
the root vectors of $A$ belonging to the eigenvalue $\lambda$,
and $\fE_{0}(A)$ (respectively $\fE(A)$) is the completion of
the vector space spanned by all the eigenvectors (respectively
root vectors) of the operator $A$. The quantities
$m_{\lambda}^{(a)}(A)=\dim\fS_{\lambda}(A)$ and
$m_{\lambda}^{(g)}(A)=\dim\Ker(A-\lambda I)$ are called
the algebraic and geometric multiplicities of $\lambda$,
respectively. By definition,  $\fE(A)\supset\fE_{0}(A)$ and
$m_{\lambda}^{(a)}\geq m_{\lambda}^{(g)}$. An eigenvalue is called
\emph{semi-simple} if $m_{\lambda}^{(a)}=m_{\lambda}^{(g)}$.
Two subspaces $\fL_{1}$ and $\fL_{2}$ of a Hilbert space are said
to be \emph{skewly linked} and denoted by $\fL_{1}\#\fL_{2}$ if
$\fL_{1}\cap\fL_{2}^{\perp}=\fL_{1}^{\perp}\cap\fL_{2}=\{0\}$.
With these preliminaries, we shall carefully reexamine the
biorthonormal relation derived in Ref.~\cite{FM81}.

The argument in Ref.~\cite{FM81} is as follows. Let $\lambda_{i}$
and $\psi_{i}$ be an eigenvalue and the corresponding eigenvector
of a non-self-adjoint operator $H$ in a Hilbert space $\fH$ equipped
with an inner product $(\cdot,\cdot)$. Then $\lambda_{i}^{\ast}$
belongs to the spectrum of $H^{\dagger}$, where $\ast$ and $\dagger$
denote complex conjugate and adjoint, respectively.
So let $\chi_{i}$ be the eigenvector of $H^{\dagger}$ corresponding
to $\lambda_{i}^{\ast}$, namely,
\begin{align}
H\psi_{i}=\lambda_{i}\psi_{i},\qquad
 H^{\dagger}\chi_{i}=\lambda_{i}^{\ast}\chi_{i}.
\label{eq:eieq}
\end{align}
Considering the relation $(\chi_{j},H\psi_{i})=(H^{\dagger}
\chi_{j},\psi_{i})$, we find $(\lambda_{i}-\lambda_{j})(\chi_{j},
\psi_{i})=0$ and thus can choose the eigenvectors such that
\begin{align}
(\chi_{j},\psi_{i})=(\psi_{i},\chi_{j})=\delta_{ij}
 \qquad\forall i,j.
\label{eq:biort}
\end{align}
Let $f$ be an arbitrary vector belonging to the space defined by
the complete set of states $\psi_{i}$ or $\chi_{i}$, namely,
$f\in\fE_{0}(A)$ or $\fE_{0}(A^{\dagger})$. We may expand
\begin{align}
f=\sum_{i}A_{i}\psi_{i}\qquad\text{or}\qquad f=\sum_{i}B_{i}\chi_{i},
\label{eq:expa}
\end{align}
where $A_{i}$, $B_{i}$ are constants. From the biorthonormality
(\ref{eq:biort}) we easily find $A_{i}=(\chi_{i},f)$ and $B_{i}=
(\psi_{i},f)$. Hence we obtain `resolutions of the identity'
\begin{align}
\sum_{i}\psi_{i}(\chi_{i},\cdot)=I\cdot\qquad\text{or}\qquad
 \sum_{i}\chi_{i}(\psi_{i},\cdot)=I\cdot{},
\label{eq:reso}
\end{align}
in terms of the biorthonormal vectors $\psi_{i}$ and $\chi_{i}$.

Although the above derivation may satisfy not a few physicists,
it cannot be justified mathematically. In fact, to justify it
rigorously, we must verify that (at least) the following conditions
are all satisfied:
\begin{description}
 \item[1] The point spectra of $A$ and $A^{\dagger}$ satisfy
  $\sigma_{p}(A^{\dagger})=\sigma_{p}(A)^{\ast}$.
 \item[2] Let $\Sigma(A)$ be the subset of $\sigma_{p}(A)$ such
  that $\Ker(A-\lambda I)\neq\Ker(A^{\dagger}-\lambda^{\ast}I)$
  for all $\lambda\in\Sigma(A)\subset\sigma_{p}(A)$. Then for all
  $\lambda\in\Sigma(A)$, each pair of eigenspaces satisfies
  $\Ker(A-\lambda I)\#\Ker(A^{\dagger}-\lambda^{\ast}I)$.
 \item[3] The geometric multiplicities satisfy $m_{\lambda}^{(g)}
  (A)=m_{\lambda^{\ast}}^{(g)}(A^{\dagger})$ for all
   $\lambda\in\Sigma(A)$ (when neither of them is finite).
\end{description}
Explanations of each condition are in order. First of all, we must
note the fact that although the spectra of $A$ and $A^{\dagger}$
always satisfy the identity $\sigma(A^{\dagger})=\sigma(A)^{\ast}$,
it does not necessarily guarantee the first condition. This is
because of the possible existence of the \emph{residual} spectrum
$\sigma_{r}(A)$ of $A$, which is defined by
\begin{align}
\sigma_{r}(A)=\bigl\{\lambda\in\bbC\bigm|\Ker(A-\lambda I)=\{0\},
 \overline{\fR(A-\lambda I)}\neq\fH\bigr\},
\end{align}
where $\fR(A)\subset\fH$ denotes the range of $A$. In general,
the relation
\begin{align}
\sigma_{r}(A)^{\ast}\subset\sigma_{p}(A^{\dagger})\subset\sigma_{r}
 (A)^{\ast}\cup\sigma_{p}(A)^{\ast}
\end{align}
follows. It is an immediate consequence of the identity
$\overline{\fR(A-\lambda I)}\oplus\Ker(A^{\dagger}
 -\lambda^{\ast}I)=\fH,$
which holds whenever $A^{\dagger}$ exists. Hence, for instance,
some eigenvalue $\lambda^{\ast}\in\sigma_{p}(A^{\dagger})$
can be related to a point of the residual spectrum
$\lambda\in\sigma_{r}(A)$ for which no corresponding
eigenvector $\psi_{i}$ exists. A similar situation can take
place between $\sigma_{p}(A)$ and $\sigma_{r}(A^{\dagger})$.
We note that in Ref.~\cite{Mo02a} the fulfillment of this first
condition is also assumed by considering only operators with a
discrete spectrum when a complete biorthonormal eigenbasis is
introduced.

Next, we recall the fact that in the case of the ordinary
orthonormality of eigenvectors $\{\psi_{i}\}$, e.g., of
a self-adjoint operator, where each eigenvector $\chi_{i}$ in
Eqs.~(\ref{eq:eieq})--(\ref{eq:reso}) is just
$\chi_{i}\propto\psi_{i}$ for all $i$, the relation (\ref{eq:biort})
for $i=j$ is guaranteed by the positive definiteness of the inner
product. In our general case $\chi_{i}\not\propto\psi_{i}$, however,
the inner product (\ref{eq:biort}) for $i=j$ can take any finite
complex number and in particular can be zero. Therefore, we cannot
choose the eigenvectors such that the relation (\ref{eq:biort})
holds without ascertaining that there exist no vectors in
$\Ker(A-\lambda I)$ (respectively in
$\Ker(A^{\dagger}-\lambda^{\ast}I)$) which is orthogonal to
$\Ker(A^{\dagger}-\lambda^{\ast}I)$ (respectively to
$\Ker(A-\lambda I)$). This is the reason why the second condition
is necessary.

Finally, the third condition must be satisfied since the relations
(\ref{eq:eieq}) and (\ref{eq:biort}) indicate the one-to-one
correspondence between the sets $\{\psi_{i}\}$ and $\{\chi_{j}\}$.
If at least one of $m_{\lambda}^{(g)}(A)$ and
$m_{\lambda^{\ast}}^{(g)}(A^{\dagger})$ is finite, this condition
is automatically satisfied under the fulfillment of the second
condition~\cite{AI89}. In this case, the second condition is also
sufficient~\cite{AI89} for the existence of a biorthonormal basis
$\{\psi_{\lambda,i},\chi_{\lambda,i}\}_{1}^{m_{\lambda}}$,
satisfying $(\chi_{\lambda,j},\psi_{\lambda,i})=\delta_{ij}$
for all $i,j=1,\dots,m_{\lambda}$, \emph{in each sector}
$\lambda\in\Sigma(A)$, where $m_{\lambda}\equiv
m_{\lambda}^{(g)}(A)=m_{\lambda^{\ast}}^{(g)}(A^{\dagger})<\infty$,
$\langle\psi_{\lambda,1},\dots,\psi_{\lambda,m_{\lambda}}\rangle
=\Ker(A-\lambda I)$, and $\langle\chi_{\lambda,1},\dots,
\chi_{\lambda,m_{\lambda}}\rangle=\Ker(A^{\dagger}-\lambda^{\ast}I)$.

We note that the above conditions are automatically satisfied in
the case of finite-dimensional spaces. In fact, the use of
a biorthonormal system is justified whenever the dimension of
the space is finite. We also note that, when
the subset $\Sigma(A)$ is empty, then we have
$\Ker(A-\lambda_{i}I)=\Ker(A^{\dagger}-\lambda_{i}^{\ast}I)$ for
all $\lambda_{i}\in\sigma_{p}(A)$ and thus can identify $\chi_{i}$
with $\psi_{i}$ for all $i$. In this case, the biorthonormal
relation (\ref{eq:biort}) reduces to just the ordinary orthonormal
one.

Another important point we should take care of, in addition to
the above conditions, is that for a non-self-adjoint operator $A$
the set of eigenvectors of either $A$ or $A^{\dagger}$ does not
generally span a dense subset of the whole Hilbert space $\fH$,
namely, $\fE_{0}(A),\fE_{0}(A^{\dagger})\subsetneqq\fH$.
Thus, even when the existence of biorthonormal system spanned
by all eigenvectors of the operator under consideration is
rigorously proved, it does not necessarily mean that the vector
$f$ which admits the expansion (\ref{eq:expa})
can be an arbitrary vector of $\fH$ and that the `resolutions of
the identity' (\ref{eq:reso}) are valid in a dense subset of $\fH$.
Therefore, we cannot apply the relation like (\ref{eq:reso})
in the whole Hilbert space unless the following additional condition
is rigorously fulfilled:
\begin{description}
 \item[4] Each set of eigenvectors of $A$ and $A^{\dagger}$ is
  complete in $\fH$, namely, $\fE_{0}(A)=\fE_{0}(A^{\dagger})=\fH$.
\end{description}
In ordinary quantum theory, it is crucial that any state vectors
in the Hilbert space $L^{2}$ can be expressed as a linear combination
of a set of the eigenstates of the Hamiltonian or physical
observables under consideration. However, this property, called
\emph{completeness}, is so frequently employed in vast areas of
applications without any doubt that one may forget the fact that
the completeness, as well as the absence of the residual spectrum,
is guaranteed by the very property of self-adjointness of
the operators.

As a simple illustration, we examine whether the above conditions
are fulfilled by a normal operator $A$, namely, an operator which
is closed, densely defined in $\fH$, and satisfies
$A^{\dagger}A=AA^{\dagger}$. In finite-dimensional spaces
the normality of operators (matrices) is the necessary and
sufficient condition for the diagonalizability by a unitary
transformation. In the general infinite-dimensional case, crucial
properties of an arbitrary normal operator $A$ in our context are
as the followings (for details, see, e.g., Refs.~\cite{AI89,BS87}):
\begin{description}
 \item[(a)] $\Ker(A-\lambda I)\perp\Ker(A-\mu I)$ for all
  $\lambda\neq\mu$.
 \item[(b)] Every eigenvalue is semi-simple.
 \item[(c)] The residual spectrum is empty, $\sigma_{r}(A)=\emptyset$,
  and thus $\sigma_{p}(A^{\dagger})=\sigma_{p}(A)^{\ast}$.
 \item[(d)] The set of eigenvectors is complete, $\fE_{0}(A)=\fE_{0}
  (A^{\dagger})=\fH$, so long as the spectrum $\sigma(A)(=\sigma
  (A^{\dagger})^{\ast})$ contains no more than a countable set of
  points of condensation.
 \item[(e)] $\Ker(A-\lambda I)=\Ker(A^{\dagger}-\lambda^{\ast}I)$
  for all $\lambda\in\sigma_{p}(A)$.
\end{description}
The first and second consequences (a) and (b) ensure
the diagonalizability of every normal operator. By virtue of
the properties (c) and (d), the conditions 1 and 4 are (almost)
always satisfied. The property (e) simply implies that the subset
$\Sigma(A)$ defined in the condition 2 is empty,
$\Sigma(A)=\emptyset$. Thus it automatically guarantees
the conditions 2 and 3, and every normal operator $A$ admits just
an ordinary orthonormal basis composed of simultaneous eigenvectors
of $A$ and $A^{\dagger}$.

Later relation (\ref{eq:reso}) was generalized to the case
when non-Hermitian Hamiltonians have block diagonal
structure~\cite{Mo02g}. It is a reasonable attempt since
non-Hermitian (more generally and adequately, non-normal)
operators in general admit non-semi-simple eigenvalues and thus
anomalous Jordan cells. In this case, it is apparent by considering
the explanations in the previous case that conditions 2--4
should be generalized to the following statements:
\begin{description}
 \item[2'] Let $\Sigma(A)$ be the subset of $\sigma_{p}(A)$ such
  that $\fS_{\lambda}(A)\neq\fS_{\lambda^{\ast}}(A^{\dagger})$
  for all $\lambda\in\Sigma(A)\subset\sigma_{p}(A)$. Then for all
  $\lambda\in\Sigma(A)$, each pair of root spaces satisfies
  $\fS_{\lambda}(A)\#\fS_{\lambda^{\ast}}(A^{\dagger})$.
 \item[3'] The algebraic multiplicities satisfy $m_{\lambda}^{(a)}
  (A)=m_{\lambda^{\ast}}^{(a)}(A^{\dagger})$ for all
   $\lambda\in\Sigma(A)$ (when neither of them is finite).
 \item[4'] Each set of root vectors of $A$ and $A^{\dagger}$ is
  complete in $\fH$, namely, $\fE(A)=\fE(A^{\dagger})=\fH$.
\end{description}
For non-normal operators acting in an infinite-dimensional Hilbert
space, however, ascertaining these conditions
2'--4' in addition to 1 is much more non-trivial and difficult.
We should again note the fact that the general situation
$\sigma_{p}(A^{\dagger})\neq\sigma_{p}(A)^{\ast}$ for non-normal
operators $A$ persists.

To conclude, the assumption of the existence of a complete
biorthonormal basis composed of eigenvectors or root vectors of
an operator puts stronger conditions on the operator. Therefore,
the results obtained under this assumption, e.g., those of
Refs.~\cite{Mo02a,Mo05b,FM81,Mo02g}, apply only to those operators
that satisfy these conditions. We hope that this work would provide
one step from naive discussions toward more fruitful and careful
investigations in and around the research field. Mathematical
characterization of a class of non-normal operators which certainly
admit a complete biorthonormal basis composed of root vectors
would be a challenging problem.

\begin{acknowledgments}
 We would like to thank A. Mostafazadeh for the valuable comment
 on the first version of this paper~\cite{Mo06b} and communication.
 We would also like to thank the anonymous referee for the
 constructive comments and suggestions.
 This work was partially supported by the National Science Council
 of the Republic of China under the grant No. NSC-93-2112-M-032-009.
\end{acknowledgments}


\bibliography{refsels}
\bibliographystyle{npb}

\end{document}